\documentclass[12pt]{iopart}

\usepackage{iopams}
\usepackage{graphicx}
\usepackage{cite}

\begin{document}

%
\title{Propagation of non-classical states of light through one-dimensional photonic lattices}

\author{B.~M. Rodr\'{\i}guez-Lara }
\address{Instituto Nacional de Astrof\'{\i}sica, \'Optica y Electr\'onica, Calle Luis Enrique Erro No. 1, Sta. Ma. Tonantzintla, Pue. CP 72840, M\'exico}
\ead{bmlara@inaoep.mx}

%
\begin{abstract}
We study the propagation of non-classical light through arrays of coupled linear photonic waveguides and introduce some sets of refractive indices and coupling parameters that provide a closed form propagator in terms of orthogonal polynomials.
We present propagation examples of non-classical states of light: single photon, coherent state, path-entangled state and two-mode squeezed vacuum impinging into two-waveguide couplers and a photonic lattice producing coherent transport.
\end{abstract}



\section{Introduction}
Classical light propagating through arrays of coupled waveguides has provided a fertile ground for the simulation of quantum physics \cite{Longhi2009p243,Longhi2011p453}.
These optical analogies of quantum phenomena are changing the way photonic integrated devices are designed; e.g. one-directional couplers \cite{DellaValle2008p011106}, light rectifiers \cite{Longhi2009p458}, isolators and polarization splitters \cite{ElGanainy2013p161105}.
As the manufacturing quality for experimental devices increases \cite{PerezLeija2013p012309}, it will soon be possible to propagate non-classical light states through linear photonic devices and a full-quantum analysis of the problem is at hand.
In quantum mechanics, propagation through an array of $N$ coupled linear waveguides is ruled by the Schr\"odinger-like equation $i \partial_{z} \vert \psi(z) \rangle = \hat{H} \vert \psi(z) \rangle$ with a Hamiltonian \cite{Bromberg2010p263604,Agarwal2012p031802},
\begin{eqnarray}
\hat{H} &=& \sum_{j=0}^{N-1} \omega_{j} \hat{n}_{j} + \sum_{j\ne k =0}^{N-2} g_{j,k} \left( \hat{a}_{j}^{\dagger} \hat{a}_{k} + \hat{a}_{j} \hat{a}_{k}^{\dagger}  \right), \label{eq:GenHam}
\end{eqnarray}
where the real parameters $\omega_{j}$ and $g_{j,k}$ are related to the effective refractive index of the $j$th waveguide and to the distance between the $j$th and $k$th waveguides, in that order. 
The operators $\hat{a}_{j}$ ($\hat{a}_{j}^{\dagger}$) annihilate (create) a photon and $\hat{n}_{j}$ gives the number of photons at the $j$th waveguide. 
Note that the vacuum state $\vert 0 \rangle$ does not couple to any other states. 
Thus we will expect that states with an important vacuum component, e.g. coherent states $\vert \alpha \rangle$ with a small coherent parameter $\vert \alpha \vert \le 1$, will serve as good examples for the peculiarities of propagation in the quantum model.

While Bloch oscillations of NOON path-entangled photons have been theoretically studied in the Heisenberg picture \cite{Bromberg2010p263604}, here we are interested in bringing forward a method in Schr\"odinger picture and introduce a class of  tight-binding waveguide arrays related to orthogonal polynomials.
Then, we study a two-waveguide coupler and propagate Fock, coherent, two-mode entangled and two-mode squeezed states. 
Finally, we analyze the propagation of such non-classical states in a well-known photonic lattice used for the coherent transport of classical fields and close with a brief discussion.

\section{Propagation in tight-binding photonic lattices}

For models with just nearest neighbor coupling, Hamiltonian (\ref{eq:GenHam}) reduces to the form:
\begin{eqnarray}
\hat{H} &=& \sum_{j=0}^{N-1} \omega_{j} \hat{n}_{j} + \sum_{j=0}^{N-2} g_{j} \left( \hat{a}_{j}^{\dagger} \hat{a}_{j+1} + \hat{a}_{j} \hat{a}_{j+1}^{\dagger}  \right). \label{eq:SpHam}
\end{eqnarray}
The equations of motion for the annihilation operators can be written in matrix form, $
\partial_{z} \vec{a}(z) = -i \mathbb{M} \vec{a}(z)$, where the auxiliary matrix $\mathbb{M}$ is tridiagonal, real and symmetric; i.e. it is a Jacobi matrix, $\left[ \mathbb{M} \right]_{j,k} = \omega_{j} \delta_{j,j} + g_{j} \left(\delta_{j,k-1} + \delta_{j-1,k} \right)$.
We have used the notation $\left[\mathbb{X}\right]_{j,k}$ for the $(j,k)$th element of matrix $\mathbb{X}$ and defined a vector of anihilation operators as $\vec{a} = \left( \hat{a}_{0},\hat{a}_{1},\hat{a}_{2}, \ldots ,\hat{a}_{N-2} ,\hat{a}_{N-1} \right)$.
It is straightforward to rewrite the matrix $\mathbb{M}$ as a product of the eigenvector matrix $\mathbb{V}$, where each row is an eigenvector, and the diagonal eigenvalue matrix $\mathbb{\Lambda}$, $\mathbb{M} = \mathbb{V}^{T} \mathbb{\Lambda} \mathbb{V}$.
The eigenvalues $\lambda_{i}$ are calculated by the method of minors as the zeros of the polynomial $p_{N}(\lambda_{i})=0$ with $p_{0}(x)=1$, $p_{1}(x)= \omega_{0} - x$, $p_{j}(x) = (\omega_{j-1} - x)p_{j-1}(x) - g_{j-2}^{2} p_{j-2}(x)$ \cite{RodriguezLara2011p053845}.
Then, we can define a set of multimode annihilation operators $\vec{A} = \mathbb{V} \vec{a}$ that diagonalize Hamiltonian (\ref{eq:SpHam}), 
\begin{eqnarray}
\hat{H} = \sum_{j=0}^{N-1} \lambda_{j} \hat{A}_{j}^{\dagger} \hat{A}_{j}.
\end{eqnarray}
For parameters that do not depend on the propagation distance, the propagator is given by:
\begin{eqnarray}
\hat{U}(z) = e^{-i z \sum_{j=0}^{N-1} \lambda_{j} \hat{A}_{j}^{\dagger} \hat{A}_{j}}.
\end{eqnarray}
Some quantities of interest that can be measured in an experimental scheme and tracked analytically are the number of photons at each waveguide, 
\begin{eqnarray}
\langle \hat{n}_{j}(z) \rangle &=& \langle \psi(0) \vert \hat{U}^{\dagger}(z) \left[\mathbb{V}^{-1} \vec{A} \right]_{j}^{\dagger} \left[\mathbb{V}^{-1} \vec{A} \right]_{j} \hat{U}(z) \vert \psi(0) \rangle,
\end{eqnarray}
where the notation $\left[ \vec{x} \right]_{j}$ has been used to represent the $j$th element of vector $\vec{x}$ and $\mathbb{X}^{-1}$ is the inverse matrix\texttt{} of $\mathbb{X}$, and the two-point correlation function,
\begin{eqnarray}
g^{(2)}_{p,q}(z) &=& \langle \psi(0) \vert \hat{U}^{\dagger}(z) \left[\mathbb{V}^{-1} \vec{A} \right]_{p}^{\dagger} \left[\mathbb{V}^{-1} \vec{A} \right]_{p} \left[\mathbb{V}^{-1} \vec{A} \right]_{q}^{\dagger} \left[\mathbb{V}^{-1} \vec{A} \right]_{q} \hat{U}(z) \vert \psi(0) \rangle,
\end{eqnarray}
that correlates the photon numbers detected at two waveguides.
We could choose a higher order detection probability $\Gamma^{(\mu,\nu)}_{p,q}=\left\langle \hat{a}^{\dagger \mu}_{p} \hat{a}^{\dagger \nu}_{q} \hat{a}^{\mu}_{p} \hat{a}^{\nu}_{q} \right\rangle$ \cite{Bromberg2010p263604}; the two-point correlation function is related to single detection $\mu = \nu =1$. 
We will also use the fidelity,
\begin{eqnarray}
\mathcal{F}(z) = \vert \langle \phi \vert \hat{U}(z) \vert \psi(0) \rangle \vert,
\end{eqnarray}
that measures how similar the propagated state $\vert \psi(z) \rangle = \hat{U}(z) \vert \psi(0) \rangle$ is to a given state $\vert  \phi \rangle$; sadly, this measurement cannot be realized experimentally without full state reconstruction.
This procedure is valid for any given set of real parameters $\{\omega_{j}, g_{j}\}$ but here we are interested in bringing forward some specific lattices related to orthogonal polynomials. 
The classical propagation of light fields has already been studied for some of these finite lattices: (i) Identical refractive indices and identical couplings \cite{SotoEguibar2011p158}, the eigenvalues are given by the roots of the $N$th Chebyshev polynomial, $U_{N}\left( \lambda_{i}/2 \right) = 0$ and the components of the eigenvector matrix $\left[\mathbb{V}\right]_{j,k} \propto U_{k}(\lambda_{j}/2)$. (ii) Identical refractive indices and couplings given by $g_{j} = g \sqrt{j+1}$ \cite{RodriguezLara2011p053845}, the eigenvalues are given by the zeros of the $N$th Hermite polynomial, $H_{N}(\lambda_{j}) = 0$, and the components of the eigenvector matrix $\left[\mathbb{V}\right]_{j,k} \propto H_{k}(\lambda_{j}/\sqrt{2})$.
(iii) Binary refractive indices, $\omega_{j} = \omega (-1)^{j}$, and identical couplings \cite{RodriguezLara2013p038116}, the eigenvalues are the roots of a Morgan-Voyce polynomial, $b_{m}(\omega^{2} - \lambda_{j}^{2})$ for even $N=2m$ and $B_{m}(\omega^{2} - \lambda_{j}^{2})$ for odd $N= 2m+1$, and $\left[\mathbb{V}\right]_{j,k} \propto b_{k/2}(\omega^{2} - \lambda_{j}^{2})$ for even $k$ and $\left[\mathbb{V}\right]_{j,k} \propto (\omega^{2} - \lambda_{j}^{2})B_{(k-1)/2}(\omega^{2} - \lambda_{j}^{2})$ for odd $k$. 
The method can also be applied for semi-infinite lattices; e.g. Refractive indices given by $\omega_{j} = (1+ \omega^{2})(j+1)$ and couplings given by $\omega \sqrt{(j+1)(j+2)}$ \cite{ZunigaSegundo2013} lead to eigenvalues $\lambda_{j}=(1-\omega^{2})(1+j)$ and to elements of the eigenvectors matrix proportional to Jacobi polynomials.


\section{Two-waveguide coupler}

In order to give a practical example, let us study propagation of non-classical light states through two coupled photonic waveguides described by the Hamiltonian \cite{Agarwal2012p031802}:
\begin{eqnarray}
\hat{H} &=& \Delta \hat{n}_{1} + g \left( \hat{a}_{1}^{\dagger} \hat{a}_{2} + \hat{a}_{1} \hat{a}_{2}^{\dagger}  \right), \\
&=& \gamma_{1} \hat{A}_{1}^{\dagger} \hat{A}_{1} + \gamma_{2} \hat{A}_{2}^{\dagger} \hat{A}_{2},
\end{eqnarray}
where the parameters $\Delta$ is a real numbers related to the refractive indices difference and the diagonalization parameters are $\hat{A}_{1} = \alpha \hat{a}_{1} + \beta \hat{\alpha}_{2}$ and $\hat{A}_{2} = \beta \hat{a}_{1} - \alpha \hat{\alpha}_{2}$ with $\alpha = 2 g \left[ 2 \Omega (\Omega-\Delta)\right]^{-1/2}$, $\beta = \left(\Omega - \Delta \right)^{1/2} \left( 2 \Omega \right)^{-1/2}$, $\Omega = \left( \Delta^{2} + 4 g^2 \right)^{1/2}$, $\gamma_{1} = \left(\Delta + \Omega \right)/2$ and $\gamma_{2} = \left(\Delta - \Omega \right)/2$ \cite{Luis1995p153}.
This Hamiltonian model is related to the quantum beam splitter described by a parameter proprotional to $gz$ \cite{Ou1987p118,Prasad1987p139,Fearn1987p485,Campos1989p1371,Leonhardt1993p3265}.
It is straightforward to calculate the photon number at each waveguide with this method, 
\begin{eqnarray}
\langle \hat{n}_{1} \rangle = \langle \psi(0) \vert \hat{n}_{1}(z) \vert \psi(0) \rangle,  \quad \langle \hat{n}_{2} \rangle = \langle  \psi(0) \vert \hat{n}_{2}(z) \vert \psi(0) \rangle, 
\end{eqnarray}
and the two-point correlation function
\begin{eqnarray}
g^{(2)}_{p,q} = \langle \psi(0) \vert \hat{n}_{p}(z) \hat{n}_{q}(z) \vert \psi(0) \rangle, \quad p,q = 1,2,
\end{eqnarray}
with $\hat{n}_{1}(z) = \left[ \Omega/ (2 g) \right]^{2} \left[ \beta^{2} \hat{A}_{1}^{\dagger} \hat{A}_{1} + \alpha^{2} \hat{A}_{2}^{\dagger} \hat{A}_{2} + g \left( e^{i \Omega t} \hat{A}_{1}^{\dagger} \hat{A}_{2} + e^{-i\Omega t} \hat{A}_{1} \hat{A}_{2}^{\dagger} \right)/\Omega \right]$ and $\hat{n}_{2}(z) = \left[ \Omega/ (2 g) \right]^{2} \left[\alpha^{2} \hat{A}_{1}^{\dagger} \hat{A}_{1} + \beta^{2} \hat{A}_{2}^{\dagger} \hat{A}_{2} - g \left( e^{i \Omega t} \hat{A}_{1}^{\dagger} \hat{A}_{2} + e^{-i\Omega t} \hat{A}_{1} \hat{A}_{2}^{\dagger} \right)/\Omega \right]$.

\begin{figure}
\centerline{\includegraphics[scale=1]{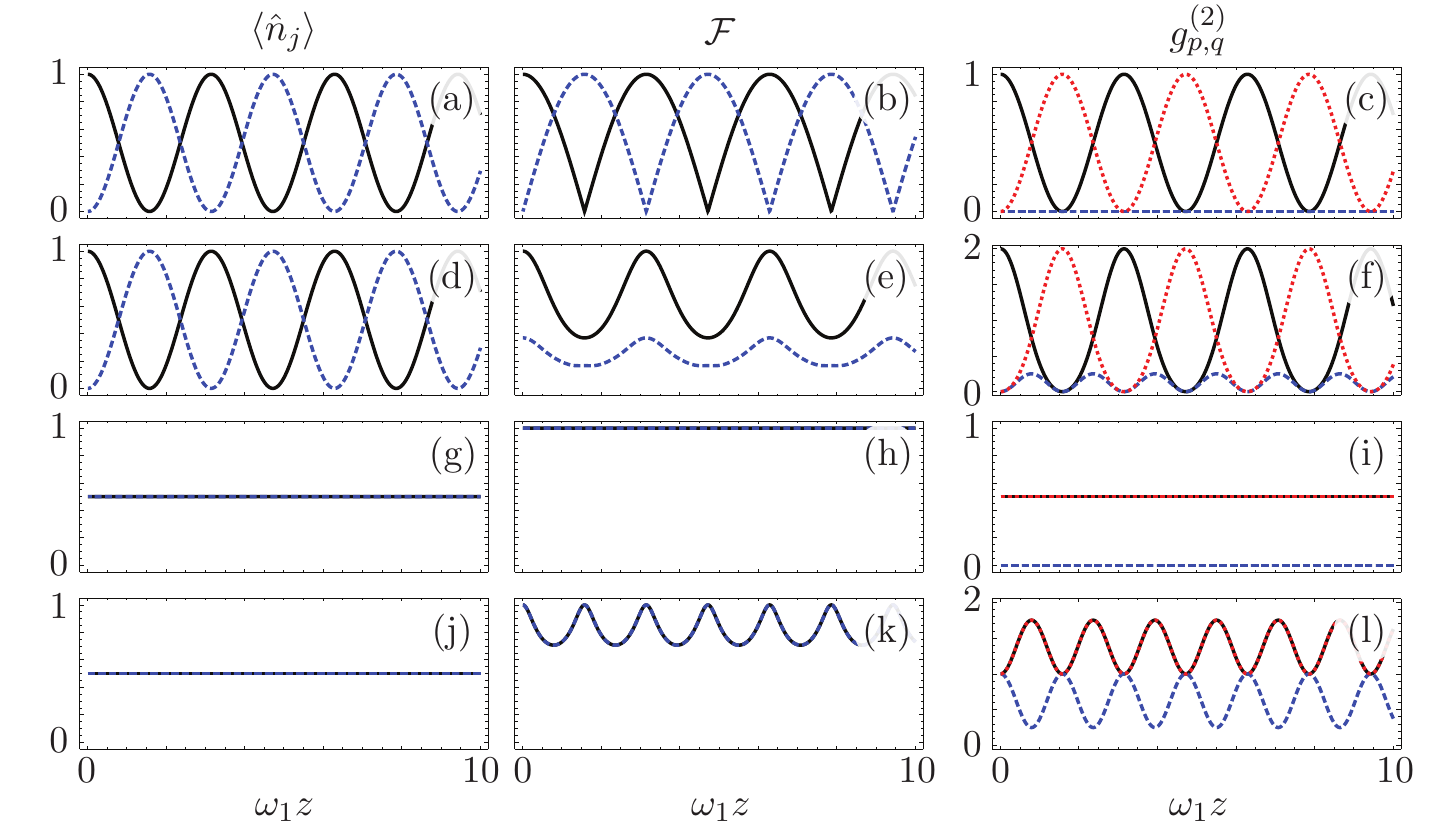}}
\caption{Propagation of the mean photon number at the first (solid black) and second (dashed blue) waveguide (first column), fidelities for finding the initial state back in the first (solid black) or transferred to the second (dashed blue) waveguide (second column) and two-point $g_{1,1}^{(2)}$ (solid black), $g_{1,2}^{(2)} = g_{2,1}^{(2)} $ (dashed blue) and $g_{2,2}^{(2)}$ (dotted red) correlation functions (third column) for a single photon $\vert \psi(0) \rangle = \vert 1,0\rangle$ (first row), a coherent state $\vert \psi(0) \rangle = \vert \alpha,0\rangle$ with $\alpha=1$ (second row), a two-mode entangled state $\vert \psi(0) \rangle = \left( \vert 1,0\rangle + \vert 0,1 \rangle \right)2^{-1/2}$ (third row) and a two-mode squeezed vacuum $\vert \psi(0) \rangle = \vert r,0\rangle$ with $r= \mathrm{arcsinh} ~2^{-1/2}$ (fourth row). The two-waveguide coupler is described by the parameter set $\Delta=0$ and $g = \omega_{1}$.}  \label{fig:Fig1}
\end{figure}

We are used to think that light will transfer from one to the other waveguide for a classical field impinging one of two-identical waveguides, $\Delta = 0$, but in the quantum case we have to remember that the vacuum state does not couple between the guides. 
Thus, whenever a non-classical state of light with a strong vacuum component impinges the two-waveguide coupler, the vacuum part of the whole state will remain at the initial waveguide; e.g. a single photon impinging one of the waveguides will transfer to the second but something different will happen for a coherent state with a mean photon number of one or less than one photon. 
For the single photon impinging the first waveguide, $\vert \psi(0) = \vert 1,0 \rangle$, the mean photon number at the waveguides are given by $\langle n_{1}(z) \rangle = \cos^{2} gt$ and $\langle n_{2}(z) \rangle = \sin^{2} gt$, Fig. \ref{fig:Fig1}(a), and the fidelity, $\mathcal{F}(z) = \beta^{2} \cos \gamma_{1} t + \alpha^{2} \cos \gamma_{2} t$, will oscillate between zero and one for identical waveguides $\Delta=0$, $\gamma_{1} = - \gamma_{2ç}$, Fig.  \ref{fig:Fig1}(b).
The two-correlation functions for identical waveguides will also show a periodic transfer of the quantum state between the waveguides, Fig. \ref{fig:Fig1}(c).
For a coherent state, $\vert \alpha \rangle = e^{- \vert \alpha \vert^2 / 2} \sum_{j=0}^{\infty} \alpha^{j} (j!)^{-1/2} \vert j \rangle$, the amplitude for the vacuum state component is proportional to $e^{-\vert \alpha\vert^2 / 2}$ and will not transfer to the second waveguide.
Thus, for a coherent state with mean photon number equal or less than one, $\vert \alpha \vert \le 1$, the vacuum amplitude will be large enough compared to the other amplitudes; e.g. for $\alpha=1$ the mean photon number at each waveguide, Fig. \ref{fig:Fig1}(d), looks identical to that of the single photon case but the fidelities, Fig.  \ref{fig:Fig1}(e), and two-point correlation functions, Fig. \ref{fig:Fig1}(f), tell us that the initial quatum state of light is never transferred to the second waveguide.
Another comparison can be made between the propagation of an initial two-mode path-entagled state $\vert \psi(0) \rangle = \left( \vert 1,0\rangle + \vert 0,1 \rangle \right)2^{-1/2}$ that has a mean photon number of one half in each waveguide and will not evolve, Fig. \ref{fig:Fig1}(g)-\ref{fig:Fig1}(i), and a two-mode squeezed state $\vert r \rangle = \mathrm{cosh}^{-1} r \sum_{j=0}^{\infty} \mathrm{tanh}^{j} r \vert j, j \rangle $ \cite{Agarwal2011p073008} which also has an average photon number of one half in each waveguide for $r=\mathrm{arcsinh 2^{-1/2}}$, Fig. \ref{fig:Fig1}(j-l).
Again the mean photon number at each waveguide for each case will look identical but the behavior of the fidelities and two-point correlation functions tell us that in the first case the initial state does not evolve while in the second case it does evolve.

\begin{figure}
\centerline{\includegraphics[scale=1]{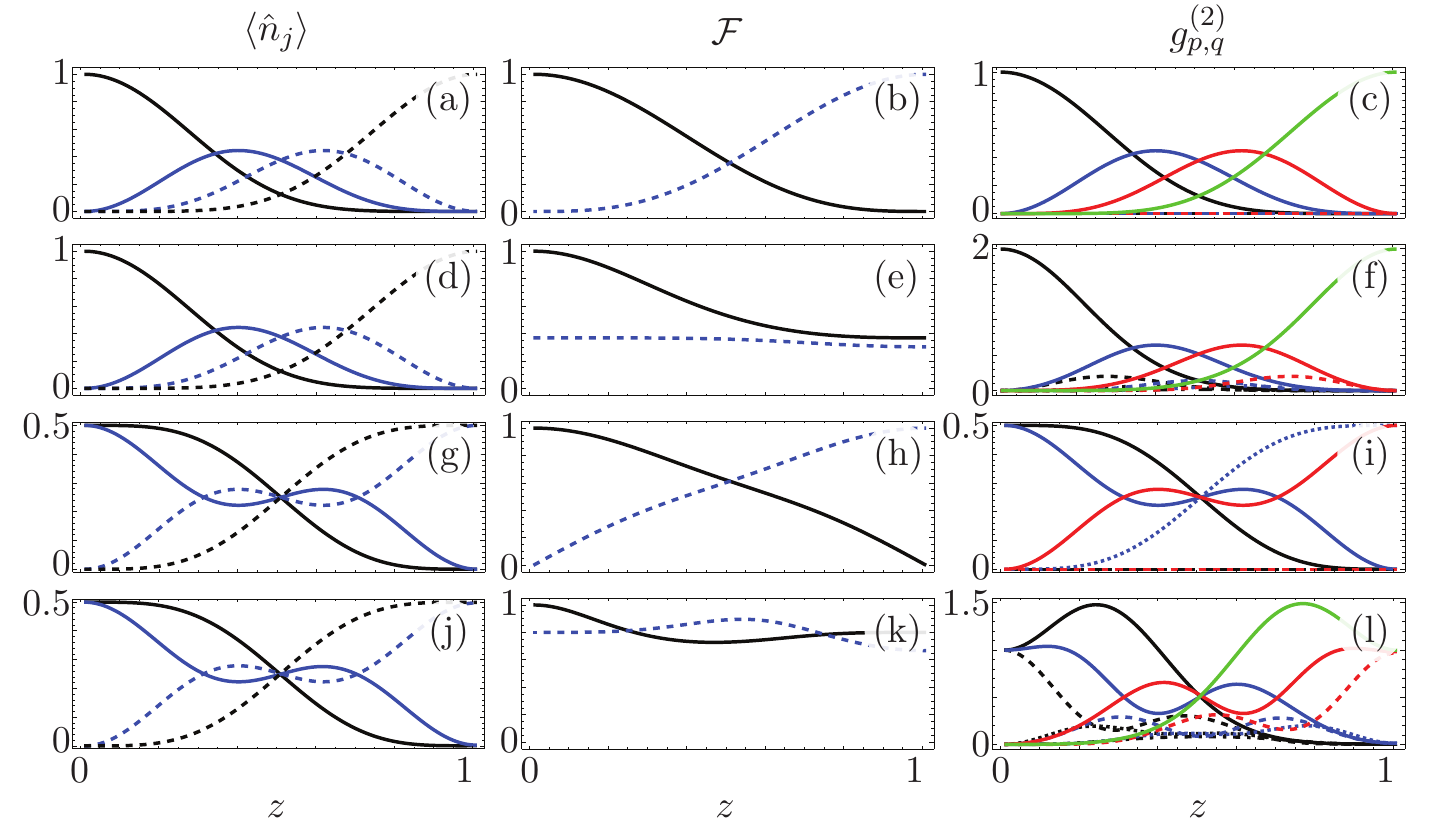}}
\caption{Propagation of the mean photon number at the first (solid black), second (solid blue), third (dashed blue) and fourth (dashed black) waveguide (first column), fidelities for finding the initial state back in the first two (solid black) or transferred to the last two (dashed blue) waveguides (second column) and two-point $g_{1,1}^{(2)}$ (solid black), $g_{1,2}^{(2)} = g_{2,1}^{(2)} $ (dashed black), $g_{1,3}^{(2)} = g_{3,1}^{(2)} $ (dotted black), $g_{1,4}^{(2)} = g_{4,1}^{(2)} $ (dot-dashed black), $g_{2,2}^{(2)}$ (solid blue), $g_{2,3}^{(2)} = g_{3,2}^{(2)} $ (dashed blue), $g_{2,4}^{(2)} = g_{4,2}^{(2)} $ (dot-dashed blue), $g_{3,3}^{(2)} $ (solid red), $g_{3,4}^{(2)} = g_{4,3}^{(2)} $ (dashed red) and $g_{4,4}^{(2)} $ (solid green) correlation functions (third column) for a single photon in the first waveguide $\vert \psi(0) \rangle = \vert 1,0,0,0\rangle$ (first row), a coherent state $\vert \psi(0) \rangle = \vert \alpha,0,0,0\rangle$ with $\alpha=1$ (second row), a two-mode entangled state $\vert \psi(0) \rangle = \left( \vert 1,0,0,0\rangle + \vert 0,1,0,0 \rangle \right)2^{-1/2}$ (third row) and a two-mode squeezed vacuum $\vert \psi(0) \rangle = \vert r,0\rangle$ with $r= \mathrm{arcsinh} ~2^{-1/2}$ (fourth row). The photonic lattice is described by the parameter set $\Delta=0$ and $g_{j} = \pi (2 z_{t})^{-1} \sqrt{j(N-j)}$ with $z_{t}=1$ and $N=4$.}  \label{fig:Fig2}
\end{figure}
\section{A photonic lattice providing perfect transfer}

Recently, a lattice providing perfect transfer of classical light described by the Hamiltonian 
\begin{eqnarray}
\hat{H} = \frac{1}{2} \sum_{j=1}^{N-1} J_{j} \left( \hat{a}^{\dagger}_{j} \hat{a}_{j+1} + \hat{a}_{j} \hat{a}^{\dagger}_{j+1} \right), \quad J_{j} = \frac{\pi}{2 z_{t}} \sqrt{j(N-j)}
\end{eqnarray}
has been proposed and demonstrated experimentally \cite{PerezLeija2013p012309,Perez-Leija2013p022303}.
The size of the lattice is given by $N$ and the parameter $z_{t}$ is the distance at which the transfer from the initial to the final end is given.
The eigenvector matrix $\mathbb{V}$ for this photonic lattice is given by Jacobi polynomials \cite{PerezLeija2013p012309} and can transfer any given state initially impinging the $j$th waveguide into the $(N-j)$th waveguide as long as the initial state does not involve a vacuum component.
When the initial state has a strong vacuum component, this particular part of the state will not transfer and this will reflect in the fidelity at the output and the two-point correlations as shown in Fig. \ref{fig:Fig2}.
Again we can see that the single photon input and the coherent state with a single photon mean photon number show identical mean photon number propagation, Fig. \ref{fig:Fig2}(a) and \ref{fig:Fig2}(d), but the fidelities, Fig. \ref{fig:Fig2}(b) and \ref{fig:Fig2}(e), and two-point correlation functions, Fig. \ref{fig:Fig2}(c) and \ref{fig:Fig2}(f), tell us that only in the single photon case the initial state is faithfully transferred to the end of the lattice.
The same happens when we compare the propagation of a two-mode path-entangled state with that of a two-mode squeezed vacuum, both with a total mean photon number of one; the mean photon number propagation at each waveguide are identical but the fidelities and two-point correlation functions will differ and point to faithful state transfer just in the case of the two-mode path-entangled state.
Figure \ref{fig:Fig2}(g)-\ref{fig:Fig2}(i) show the propagation results for the two-mode entangeld state and Fig.\ref{fig:Fig2}(j)-\ref{fig:Fig2}(l) show the propagation of a two-mode squeezed vacuum state with a squeezing parameter $r= \mathrm{arcsinh} ~2^{-1/2}$ leading to a total mean photon number of one.

\section{Conclusion}

We have shown a method to deal with non-classical light propagating through tight-binding arrays of coupled linear waveguides. 
While we focused on listing refractive indices and coupling parameters that allow us to diagonalize the quantum Hamiltonian by use of multimode annihilation operators whose cofficients are given in terms of orthogonal polynomials, the method can be used for any given set of parameters as a real symmetric tridiagonal matrix is always feasible of diagonalization.
For the sake of bringing forward one of the most pronounced difference between propagation of classical and non-classical states of light, we focused in comparing the propagation of states lacking a vacuum component, single-photon and a two-mode entangled state, with those showing a significant vacuum component, coherent and two-mode squeezed state with one or less than one mean photon number.

\section*{Acknowledgments}
The author is grateful to G. S. Agarwal for suggesting the study of two-mode squeezed state propagation in coupled waveguides.


\section*{References}
\providecommand{\newblock}{}

\end{document}